\documentclass[aps,prl,twocolumn,superscriptaddress]{revtex4-1}
\usepackage{graphicx}
\usepackage{dcolumn}
\usepackage{bm}
\usepackage{amsmath}
\usepackage{amssymb}
\usepackage{latexsym}
\usepackage{epsfig}
\usepackage{amsbsy}
\usepackage{array}
\usepackage{amssymb}
\usepackage{setspace}
\usepackage{float,url}
\usepackage{textcomp}
\usepackage{multirow}
\usepackage{threeparttable}
\usepackage{epstopdf}
\usepackage[colorlinks=true,
                        linkcolor=blue,
                        anchorcolor=blue,
	                     urlcolor=blue,
                        citecolor=blue]{hyperref}

\begin{document}

\title{On the Nonlinear Excitation of Phononic Frequency Combs in Molecules}

\author{Hongbin Lei}
\affiliation{\mbox{Department of Physics, National University of Defense Technology, Changsha 410073, China}}
\author{Qian Zhang}
\affiliation{\mbox{Department of Physics, National University of Defense Technology, Changsha 410073, China}}
\author{Hongqiang Xie}
\affiliation{\mbox{School of Science, East China University of Technology, Nanchang 330013, China}}
\author{Congsen Meng}
\affiliation{\mbox{Department of Physics, National University of Defense Technology, Changsha 410073, China}}
\author{Zhaoyang Peng}
\affiliation{\mbox{Department of Physics, National University of Defense Technology, Changsha 410073, China}}
\author{Jinlei Liu}
\affiliation{\mbox{Department of Physics, National University of Defense Technology, Changsha 410073, China}}
\author{Guangru Bai}
\affiliation{\mbox{Department of Physics, National University of Defense Technology, Changsha 410073, China}}
\author{Adarsh Ganesan}
\email{adarsh@dubai.bits-pilani.ac.in}
\affiliation{\mbox{Department of Electrical and Electronics Engineering, Birla Insitute of Technology and Science (BITS), Pilani - Dubai Campus} \mbox{Dubai 345055, UAE}}
\affiliation{\mbox{Department of Mechanical Engineering, Birla Insitute of Technology and Science (BITS), Pilani - Pilani Campus} \mbox{Pilani 333031, India}}
\author{Zengxiu Zhao}
\email{zhao.zengxiu@gmail.com}
\affiliation{\mbox{Department of Physics, National University of Defense Technology, Changsha 410073, China}}

\renewcommand{\thefootnote}{\fnsymbol{footnote}}

\footnotetext[2]{You can add acknowledgements here.}
\date{\today}

\begin{abstract}
The mechanical analog of optical frequency combs, phononic frequency combs (PFCs), has recently been demonstrated in mechanical resonators via nonlinear coupling among multiple phonon modes. However, for exciting phononic combs in molecules, the requisite strong nonlinear couplings need not be readily present. To overcome this limitation, this paper introduces an alternative route for the generation of phononic combs in polar molecules. Theoretically, we investigated the radiation and phononic spectra generated from CO molecule possessing relatively large permanent dipole moment with density matrix formalism. By considering rovibronic excitation of the ground-state CO molecule while avoiding the electronic excitation, the contribution of the permanent dipole moment and electric dipole polarizability to the creation of PFCs is demonstrated and distinguished. The finding could motivate the possible extension of combs to molecular systems to offer new avenues in molecular sciences.
\end{abstract}


\maketitle

Phononic frequency combs (PFCs) are a newly documented phenomenon in the domains of phononics, mechanics, vibrations and acoustics. The spectral features of phononic combs are characterized by an array of equidistant frequencies. The first demonstration of this phenomenon was conducted using a mechanical resonator \cite{GanesanPRL2017}. The paper \cite{GanesanPRL2017} shows that the coupling between length extensional and flexural vibration modes aid in the generation of frequency combs. The follow-on papers \cite{GanesanAPL2018,GanesanPRB2018} showcased different dynamical pathways that lead to the observations of phononic combs. While we used a piezoelectric micromechanical beam as the experimental platform for all these studies \cite{GanesanPRL2017,GanesanAPL2018,GanesanPRB2018}, other research groups extended the results of phononic combs to other platforms including a coupled translational-rotational resonator \cite{Czaplewski2018}, a bulk-acoustic wave resonator \cite{Goryachev2020}, a graphene resonator \cite{Singh2020}, a MoS$_2$ resonator \cite{chiout2021multi} and gas bubble clusters \cite{nguyen2021acoustic}. Despite this growing number of experimental observations of PFCs in mechanical resonators, the existence conditions for such combs was largely unknown. Qi et al. presented the analytical expressions describing the existence conditions for phononic frequency combs  \cite{Qi2020}.

The applicability of these analytics is to any vibratory system. However, since the generation mechanism involves nonlinear coupling among multiple modes, the utilization of this pathway for exciting frequency combs in molecules may not be straightforward. Inspired by the recent observations of PFC in a linear mechanical resonator using optical tweezers \cite{de2023mechanical}, this paper defines vibrational high-harmonic generation as a potential comb excitation mechanism for molecules. The PFCs from molecules we discussed here originate from the energy transfer of the light by photon absorption.  In order to effectively excite the vibration motion while keeping the electron in the electronic ground state, molecules with permanent dipole moments are favorite candidates. Note that photon momentum is ignored here due to the large wavelength of the driving pulse. For mechanical resonators, the PFCs are excited by mechanical forces which transfer momentum to the media and excite the PFCs. In a different setup using tightly focusing light like optical tweezer where photon momentum is important, the excitation of PFCs from molecules will be similar to mechanical resonator \cite{de2023mechanical}.

In the present work, we choose the polar CO molecule as the target, and perform a proof-of-principle demonstration that the PFCs can be effectively generated when the CO molecules are ro-vibrationally excited by a weak mid-infrared femtosecond laser. With the time-dependent density matrix theory, the physical mechanism for the generation of radiation and phonon spectra from the polar molecules is uncovered by considering two situations of one-photon and two-photon resonances. Our simulations indicate that both the permanent dipole moment and electric dipole polarizability are crucial for the creation of the radiation and phonon spectra, and the current approach can be generalized to the relevant studies of PFCs from other molecules.

\section{II. Theory}
\subsection{A. Density Matrix}
In our model,  a long-wavelength and linearly polarized pump laser with relatively low intensity is employed to effectively excite the rovibrational modes of a linear polar molecule while the electronic excitation is avoided. The time-dependent Schr$\rm{\ddot{o}}$dinger equation (in atomic units) describing the rovibrational excitation dynamics under the Born-oppenheimer approximation can be expressed as
\begin{equation}\label{eq1}
i\frac{\partial}{\partial t}\Psi(\bm{R},t)=\mathrm{H}(t)\Psi(\bm{R},t),
\end{equation}
where $\bm{R}=(R,\theta,\phi)$ is the internuclear distance vector and the Hamiltonian $\mathrm{H}(t)$ of single electronic states can be written as \cite{Aubanel1994,Friedrich1995,Hoang2017},
\begin{align}
&\mathrm{H}=\mathrm{H_{0}}+\mathrm{H_{1}}+\mathrm{H_{2}},\label{eq2}\\
&\mathrm{H_{0}}=-\frac{1}{2M}\left(\frac{\partial^2}{\partial R^2}-\frac{{\hat{J}}^2}{R^2}\right)+V(R),\label{eq3}\\
&\mathrm{H_{1}}=-\mu(R)\varepsilon(t)\cos\theta,\label{eq4}\\
&\mathrm{H_{2}}=-\frac{1}{2}\varepsilon^2(t)(\alpha_\parallel(R)\cos^2\theta+\alpha_\bot(R)\sin^2\theta),
\end{align}
where $\mathrm{H_{0}}$ is the free-field Hamiltonian, $M$ is the reduced mass of the two nuclei, $\hat{J}$ is the total angular momentum operator of the molecule, $V(R)$ is the internuclear interaction as a function of the internuclear distance $R$, i.e., the potential energy curve, $\mathrm{H_{1}}$ corresponds to the interaction of the permanent dipole moment $\mu(R)$, $\varepsilon(t)$ is the electric field of pump laser, and $\theta$ is the angle between the polarization direction of pump laser and the molecular axis. In the last term of Eq. (\ref{eq2}), we take into account the interaction of the electric dipole polarizability describing the charge anisotropic polarization of the electronic state within the laser pulse, which is given by the two components parallel and perpendicular to the molecular axis, i.e., the polarizabilities $\alpha_{\parallel}(R)$ and $\alpha_{\bot}(R)$ as a function of $R$.

The time-dependent molecular wave function $\Psi(\bm{R},t)$ can be expanded as
\begin{align}\label{eq6}
&\Psi(\bm{R},t)=\sum_{\nu, J, m}C_{\nu J m}(t)\Phi_{\nu J m}(R,\theta,\phi),\\
&\Phi_{\nu J m}(R,\theta,\phi)=\psi_{\nu J}(R)\mathrm{Y}_{J m}(\theta,\phi),
\end{align}
where $C_{\nu J m}(t)$ is the probability amplitude of the eigenstate $\Phi_{\nu Jm}(R,\theta,\phi)$ with the vibrational quantum number $\nu$, the rotational quanmtum number $J$ and the projection $m$ of rotational quanmtum number $J$ on the laser polarization direction. The eigenstate $\Phi_{\nu Jm}(R,\theta,\phi)$ consists of the radial wave functions $\psi_{\nu J}(R)$ and the spherical harmonics $\mathrm{Y}_{Jm}(\theta,\phi)$ and satisfies eigenvalue equation,
\begin{equation}\label{eq8}
\begin{aligned}
\mathrm{H_{0}}(t)|\nu Jm\rangle=E_{\nu J}|\nu Jm\rangle,
\end{aligned}
\end{equation}
where $\Phi_{\nu Jm}(R,\theta,\phi)$ is abbreviated by the Dirac notation $|\nu Jm\rangle$ and $E_{\nu J}$ is the eigenvalue corresponding to the rovibronic states $|\nu Jm\rangle$ with degeneracy of $2J+1$.

For simulating the nonlinear rovibrational excitation of linear polar molecule within the laser pulse under the initial condition of mixed quantum ensemble, the density matrix of the quantum ensemble can be defined as
\begin{equation}\label{eq9}
\begin{aligned}
\hat{\rho}(t)=|\Psi(\bm{R},t)\rangle\langle\Psi(\bm{R},t)|.
\end{aligned}
\end{equation}
The density matrix is equivalent to the wave function and contains all the dynamics of molecular quantum system. The evolution of density matrix over time can be described with the Liouville-von Neumann equation \cite{Blum2012} (in atomic units),
\begin{equation}\label{eq10}
\begin{aligned}
\frac{\partial}{\partial t}\hat{\rho}(t)=-i\left[\mathrm{\hat{H}}(t),\hat{\rho}(t)\right],
\end{aligned}
\end{equation}
where $\mathrm{\hat{H}}$ is the Hamiltonian matrix. Based on the Eqs. (\ref{eq1})(\ref{eq2}), the Hamiltonian matrix $\mathrm{\hat{H}}$ can be written as
\begin{align}\label{eq11}
\mathrm{\hat{H}}&=\mathrm{\hat{H}_{0}}+\mathrm{\hat{H}_{1}}+\mathrm{\hat{H}_{2}}.
\end{align}
Here, $\mathrm{\hat{H}_{0}}$ is the diagonal matrix of eigenenergies $E_{\nu J}$. $\mathrm{\hat{H}_{1}}$ and $\mathrm{\hat{H}_{2}}$ are the interaction matrix from the permanent electric dipole moment and the electric dipole polarizability, respectively. These two interaction matrixes can be written as
\begin{equation}
\begin{aligned}
\mathrm{\hat{H}_{1}}=-\varepsilon(t)\hat{d}_{1};\mathrm{\hat{H}_{2}}=-\frac{1}{2}\varepsilon^{2}(t)\hat{d}_{2},\label{eq12}
\end{aligned}
\end{equation}
where $\hat{d}_{1}$ and $\hat{d}_{2}$ with the elements $d_{1}^{\nu Jm,\nu'J'm'}$ and $d_{2}^{\nu Jm,\nu'J'm'}$ are the transition matrixes of permanent electric dipole and electric dipole polarizability, respectively. The elements $d_{1}^{\nu Jm,\nu'J'm'}$ and $d_{2}^{\nu Jm,\nu'J'm'}$ can be expressed as 
\begin{align}
	&d_{1}^{\nu Jm,\nu' J'm'}=\langle\nu Jm|\mu(R)\cos\theta|\nu' J'm'\rangle,\label{eq13}\\
	&d_{2}^{\nu Jm,\nu' J'm'}=\langle\nu Jm|\alpha_\parallel(R)\cos^2\theta+\alpha_\bot(R)\sin^2\theta|\nu' J'm'\rangle.\label{eq14}
\end{align}

It can be known from Eq.(\ref{eq12}) that $\mathrm{\hat{H}_{1}}$ and $\mathrm{\hat{H}_{2}}$ are proportional to the laser electric field and the square of laser electric field, respectively. Thus, $\mathrm{\hat{H}_{1}}$ and $\mathrm{\hat{H}_{2}}$ can induce the rovibrational excitation by absorbing an odd and even number of photons, respectively, while the electronic excitation is screened. In addition, we can expand the permanent dipole moment $\mu(R)$ and the electric dipole polarizability $\alpha_{\parallel,\bot}(R)$ around the equilibrium internuclear distance $R_e$ to analyze the vibrational excitation of polar molecules induced by them within the pump laser. The permanent dipole moment can expended as, $\mu(R)=\mu_0(R_e)+\frac{d\mu}{dR}(R-R_e)+\frac{1}{2}\frac{d^2\mu}{dR^2}{(R-R_e)}^2+\cdots$. Here, the intrinsic dipole moment $\mu_0(R_e)$ governs the rotational excitation on the same vibrational level, the first order derivative $\frac{d\mu}{dR}$ governs the vibrational excitation with selection rule $\Delta\nu=\pm1$ and the additional terms will govern vibrational transition selection rule $\Delta\nu=\pm2, \pm3, \cdots$, where $\Delta\nu=\nu'-\nu$ marking the vibrational quantum number of upper and lower levels as $\nu'$ and $\nu$, respectively. As same as the permanent dipole moment, the electric dipole polarizability can also be expended around the equilibrium internuclear distance $R_e$ as $\alpha_{\parallel,\bot}(R)=\alpha_{\parallel,\bot}^{0}(R_e)+\frac{d\alpha_{\parallel,\bot}}{dR}
(R-R_e)+\frac{1}{2}\frac{d^2\alpha_{\parallel,\bot}}{dR^2}{(R-R_e)}^2+\cdots$, each term of which governs vibrational excitation with selection rule similar to that of permanent dipole moment. Furthermore, the selection rule of rotational excitation of polar molecule  can be obtained by the properties of spherical harmonics. Since the interaction term between linear polarization and molecular system is independent of the precession angle $\phi$, the projection $m$ of rotational quantum number $J$ remains constant during rovibrational excitation, i.e., $\Delta m=0$. The angular term $\cos\theta$ of $\mathrm{\hat{H}_{1}}$ determines the rotational excitation with selection rule, $\Delta J=J'-J=\pm1$, marking the total angular momentum of upper and lower levels as $J'$ and $J$, respectively. Differing from $\mathrm{\hat{H}_{1}}$, the angular term $\cos^2\theta(\sin^2\theta)$ of $\mathrm{\hat{H}_{2}}$ governs the rotational excitation with selection rule, $\Delta J=0,\pm2$. Generally, it is usual to describe rotational transitions with $\Delta J=-2, -1, 0, 1, 2$, as O-branch, P-branch, Q-branch, R-branch, S-branch transitions \cite{Seideman2005,Bransden1982,Brand1960}, respectively.
\begin{figure}
\includegraphics*[width=3.4in]{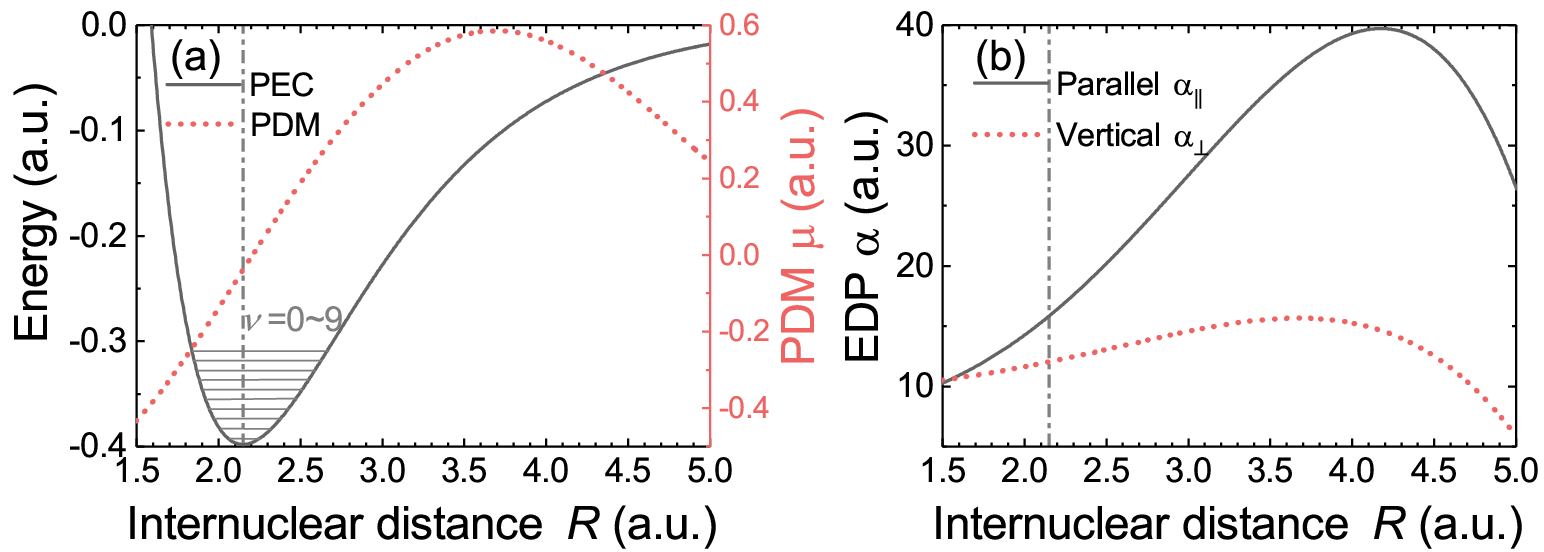}
\setlength{\abovecaptionskip}{-5pt}
\setlength{\belowcaptionskip}{0pt}
\caption{
(a) The potential energy (black solid line) and the permanent dipole moment (red dashed line) as a function of the internuclear distance $R$ for CO molecule staying in the ground electronic state $X{^1}\Sigma^+$. The eigenenergies of vibrational levels $\nu=0\sim9$ are indicated by grey solid lines. (b) The black solid and red dashed lines are parallel and perpendiculvar components of electric dipole polarizability as a function of the internuclear distance $R$, respectively. The chain-dotted line in (a)(b) indicates the equilibrium position $R_{e}=2.137$ a.u..
}\label{f1}
\end{figure}

In this work, we pick the CO molecule as an example since CO molecule is a highly polar molecule and has both the permanent dipole moment and electric dipole polarizability, which vary greatly with the internuclear distance $R$. Figure \ref{f1}(a) shows the potential energy curve of the ground electronic state of CO molecule \cite{Borges2001} and its permanent dipole moment (PDM) varying with the internuclear distance \cite{Langhoff1995}. The positive value of PDM correspond to that PDM points from C to O, i.e., C$^+$O$^-$ polarity. It can be seen from Fig. \ref{f1}(a) that the potential energy curve is asymmetric parabolic-like shape, resulting in the detuning of energy differences between adjacent vibrational states. The parallel component $\alpha_{\parallel}(R)$ and vertical component $\alpha_{\bot}(R)$ of the electric dipole polarizability (EDP) \cite{Maroulis1996} as a function of $R$ are exhibited in the Fig. \ref{f1}(b). In addition, the equilibrium internuclear distance $R_e$ is 2.137 a.u. indicated by the chain-dotted line, where the permanent dipole moment and electric dipole polarizability have a large variation with the internuclear distance. Note that the rovibrational excitation is determined by the derivatives of the interaction with the permanent dipole moment and the electric dipole polarizability instead of the dipole interactions themselves, as noted above.
For CO molecule, the derivatives at the equilibrium internuclear distance are $\mu^\prime\left(R_e\right)\approx0.67$ a.u., $\alpha_\parallel^\prime\left(R_e\right)\approx11.3$ a.u., and $\alpha_\bot^\prime\left(R_e\right)\approx2.7$ a.u., respectively \cite{Sunil1988}. Consequently, when the CO molecule is driven by the laser field to vibrate, the  molecular internuclear distance is modulated and the molecule is stimulated to the excited vibrational levels.

\begin{figure*}[htbp]
\includegraphics*[width=7.1 in]{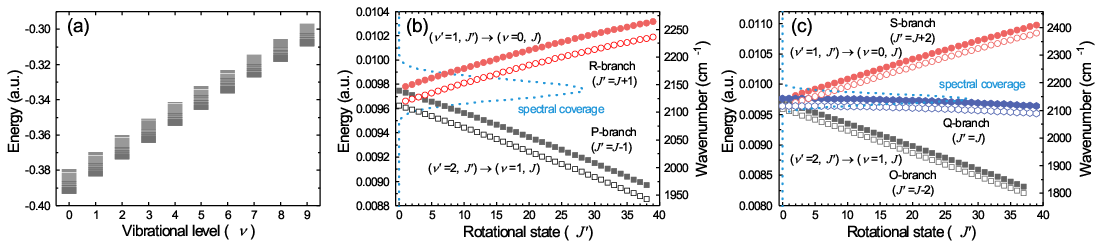}
\setlength{\abovecaptionskip}{-5pt}
\setlength{\belowcaptionskip}{0pt}
\caption{
(a) The eigenergy distribution $E_{\nu, J}$ of rotational state on the vibrational level. (b) The R-branch and P-branch transition energy of CO molecule as a function of rotational quantum number $J'$ is depicted with red circle and black square, respectively. (c) The S-branch, Q-branch and O-branch transition energy of CO molecule as a function of rotational quantum number $J'$ is exihibited by red circle, blue diamond and black square, respectively. In (b) and (c), the solid and hollow dots indicate the transition from $(\nu'=1, J')$ to $(\nu=0, J)$ and from $(\nu'=2, J')$ to $(\nu=1, J)$, respectively. The light blue dotted line represents the spectral coverage of employed pulse.
}\label{f2}
\end{figure*}

Furthermore, the transition wavelength from the ground electronic state ($X{^1}\Sigma^+$) to the first excited state ($a{^3}\prod_\mathrm{r}$) is 205.4 nm for the CO molecule \cite{Helene1961}. Therefore, we choose the pump wavelength in several $\mathrm{\mu}$m whose corresponding photon energy is far less for the electronic excitation by absorbing one photon but comparable to the vibrational excitation frequency, i.e., $\omega_{v}\sim\omega_{l}\ll\omega_{e}$, where $\omega_\nu$ is the transition frequency of vibrational states of the ground electronic state, $\omega_{l}$ is the pump laser frequency, and $\omega_e$ is the electronic transition frequency. For the sake of choosing the proper pump wavelength to excite the vibrational motion of CO molecule, the rovibrational eigenstates $\Phi_{\nu J m}(R,\theta,\phi)$ and eigenergies $E_{\nu J}$ are calculated by Eq. (\ref{eq3}) and Eq. (\ref{eq8}), as displayed in Fig. \ref{f2}(a). The transition energy of vibrational excitation with $\Delta\nu=\pm1$ as a function of a rotational quantum number $J'$ is shown in Figs. \ref{f2}(b)(c). Figures \ref{f2}(b) and (c) clearly show that the transition energies of rovibrational excitation from $(\nu'=1,$ $J')$ to $(\nu=0,$ $J)$ are around $\omega_{v}=0.00975$ a.u. and the transition energy decreases with the vibrational quantum number increasing due to the asymmetry of the potential energy curve, e.g., the transition energy from $(\nu'=2$, $J')$ to $(\nu=1,$ $J)$ is lower than that of transition from $(\nu'=1,$ $J')$ to $(\nu=0,$ $J)$. In the simulation, we employ the driving laser pulse with a Gaussian envelope $\varepsilon(t)=\varepsilon_{0}\exp(-t^2/\tau^2)\cos(\omega_{l} t)$ to excite CO molecule to vibrate. The laser center frequency is chosen to be $\omega_{l}=0.00975$ a.u. (4.672 $\mathrm{\mu}$m) or $\omega_{l}=0.004875$ (9.344 $\mathrm{\mu}$m) a.u. for exciting the vibration excitation of CO molecule via the one-photon or two-photon resonance channels, respectively. Moreover, the spectral coverage of employed laser pulse with the pulse duration of $\tau =500fs$ and the center frequency of $\omega_{l}=0.00975$ a.u. is depicted in Figs. \ref{f2}(b)(c) by the light blue dotted line. Obviously, the employed spectra can cover many transition energies of low energy rotational states.

In the meantime, the system of CO molecule is in thermal equilibrium at initial time, which means that the population distribution of CO molecule obeys the Boltzmann distribution \cite{Pinkham2005},
\begin{equation}\label{eq15}
\begin{aligned}
P_{\nu, J}=\frac{2J+1}{A}\exp\left(-\frac{E_{\nu J}}{k_{B}\mathrm{T}}\right),
\end{aligned}
\end{equation}
where $A$ is normalization coefficient, $k_{B}$ is the Boltzmann constant and $\mathrm{T}=293.15$ K is the room temperature. The initial population distribution of rotational states for different projections $m$ of the same $J$ are uniform because they are degenerate states. The initial population distribution of CO molecule can be obtained by substituting $E_{\nu J}$ into Eq. (\ref{eq15}), which can be found that CO molecules are initially on the vibration ground state $\nu=0$ and the rotational state of the largest population is $J=7$. Therefore, the pulse used in our simulation with spectral width covering the many transition energies of low energy rotational states is sufficient to excite the rovibrational motion of CO molecule.

\begin{figure}
\includegraphics*[width=3.4in]{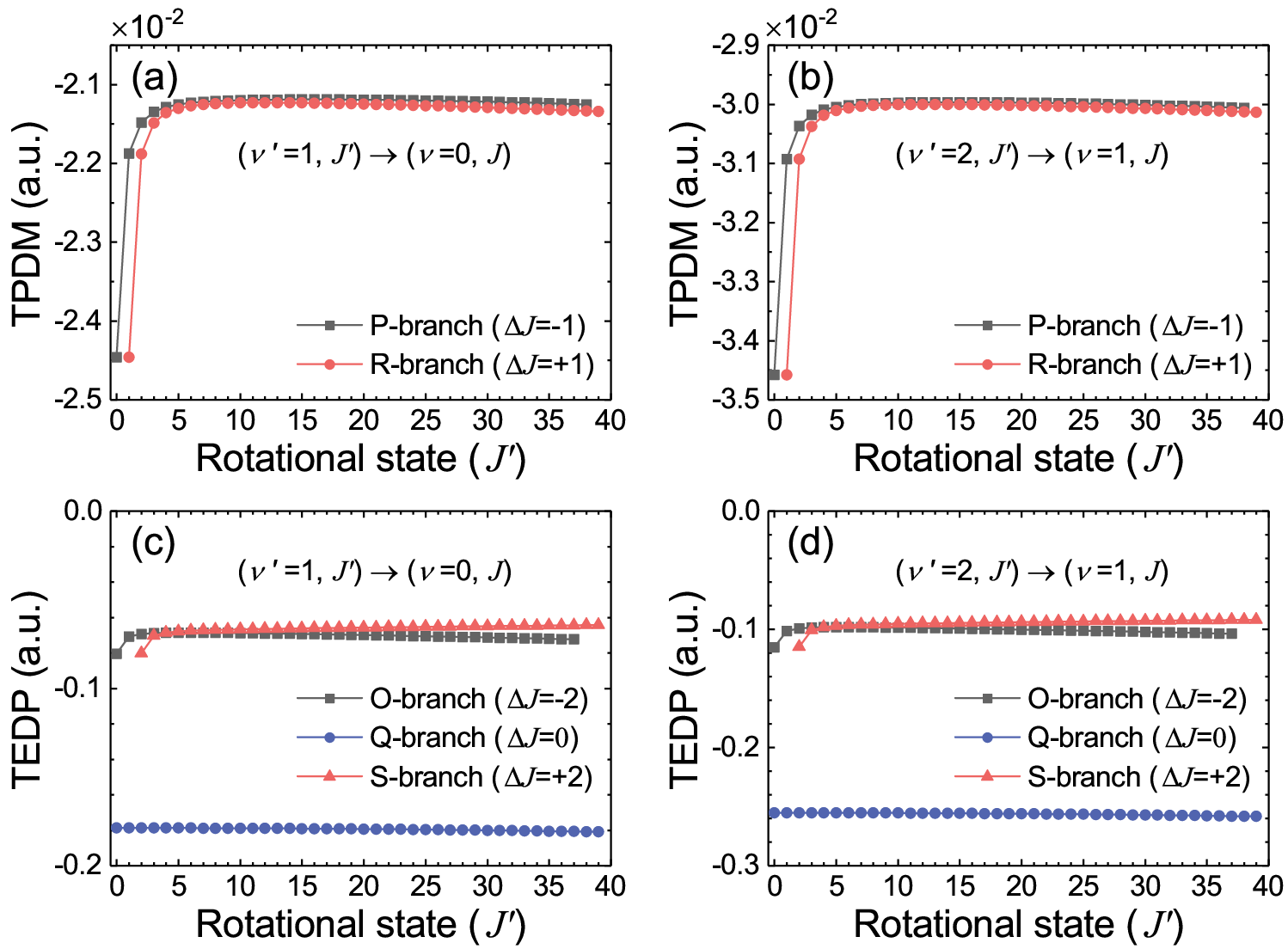}
\setlength{\abovecaptionskip}{-2pt}
\setlength{\belowcaptionskip}{0pt}
\caption{
Transition matrix elements of permanent dipole moment (TPDM) and electric dipole polarizability (TEDP) as a function of rotational quantum number $J'$ is shown in (a)(b) and (c)(d), respectively. (a)(c) and (b)(d) represent the transition moment from $(\nu'=1, J')$ to $(\nu=0, J)$ and $(\nu'=2, J')$ to $(\nu=1, J)$, respectively. In (a) and (b), the black square and red circle represent the transition of P-branch $(\Delta J=-1)$ and R-branch $(\Delta J=+1)$, respectively. In (c) and (d), the black square, blue diamond, and red circle represent the O-branch $(\Delta J=-2)$, Q-branch $(\Delta J=0)$, and S-branch $(\Delta J=+2)$ transitions, respectively. The projection $m'$ of the rotational quantum number $J'$ on the laser polarization direction is chosen to be $m'=m=0$ and $\Delta J=J'-J$.
}\label{f3}
\end{figure}

Moreover, the transition matrix elements are calculated by substituting the rovibrational eigenstates $\Phi_{\nu J m}(R,\theta,\phi)$ into Eq. (\ref{eq13}) and (\ref{eq14}). In order to compare the role of electric dipole moment and electric dipole polarizability in driving the vibrations of CO molecules, transition matrix elements of permanent dipole moment (TPDM) and electric dipole polarizability (TEDP) as a function of rotational quantum $J'$ with $m'=m=0$ are presented in Fig. \ref{f3}. As illustrated in Fig. \ref{f3}, the transition matrix elements gradually tend to a constant with the increase of the rotational quantum number $J'$\cite{zhangyouyuan2020}. Besides, when the peak laser intensity is given by 5$\times$10$^{13}$ W/cm$^2$ (I$_\mathrm{0}=\varepsilon_{0}^{2}=$0.00142 a.u.), the transition probability induced by the permanent electric dipole and electric dipole polarizability are in the same order of magnitude, i.e., $\varepsilon_{0}\left|d_{1}^{\nu Jm,\nu'J'm'}\right|\sim \mathrm{I_{0}}\left|d_{2}^{\nu Jm,\nu'J'm'}\right|$ \cite{Lewis2000}. Thus, the electric dipole polarizability is as important as the electric permanent dipole moment in driving the rovibrational excitation of CO molecules under our consideration, and which of these dominants the rovibrational excitation depends on the pump wavelength. 

\subsection{B. Radiation and phononic spectra}
The electric dipole radiation has been researched extensively, but the radiation induced by the electric dipole polarizability is rarely studied. Next, we will derive the formula for calculating the radiation spectra including the permanent electric dipole moment and electric dipole polarizability, as well as give the formula to calculate the phononic spectra.

For electric charges in space, the induced vector potential $\bm{A}(\bm{r})$ at the spatial position $\bm{r}$ with Taylor expansion is given by \cite{Jackson1998}
\begin{equation}\label{eq16}
\begin{aligned}
\bm{A}(\bm{r})=\frac{\mu_{0}}{4\pi x}\int_{V}\bm{J}(\bm{r'})(1-ik\bm{e_{x}}\cdot\bm{r'}+\cdots)dV',
\end{aligned}
\end{equation}
where $\mu_{0}$ is the permeability of free space, $\bm{J}(\bm{r'})$ is the electric current density and $x$ is the distance from $\bm{r}$ to the electric charges with the radial direction unit vector $\bm{e_{x}}$. By neglecting the high-order terms and the magnetic interaction, $\bm{A}(\bm{r})$ can be reduced,
\begin{equation}\label{eq17}
\begin{aligned}
\bm{A}(\bm{r})=\frac{\dot{\bm{D}}e^{ikx}}{4\pi\epsilon_{0}c^{2}x}
\end{aligned}
\end{equation}
where $\epsilon_{0}$ is the permittivity of free space, $k=\omega/c$, $\omega$ is the radiation frequency, $c$ is the light speed, $\bm{D}$ is the electric dipole moment vector.

Using the Maxwell's equations and the substitution formula, i.e., $\nabla\leftrightarrow ik\bm{e_{x}}$ and $\partial/\partial t\leftrightarrow -i\omega$, the electric field induced by the dipole moment can be written as
\begin{align}\label{eq18}
\bm{E}&=\frac{ic}{k}\nabla\times\nabla\times\bm{A}=\frac{\ddot{D}e^{ikx}\sin\theta}{4\pi\epsilon_{0}c^{2}x}\bm{e_{\theta}},
\end{align}
where $\bm{e_{\theta}}$ indicates the direction of generated electric field.

Based on the theory of molecular polarizabilities \cite{McLean1967}, the laser induced molecular dipole moment $D(t)$ can be expressed as
\begin{equation}\label{eq18e}
	\begin{aligned}
		D(t)=\mu(t)+\alpha(t)\varepsilon(t).
	\end{aligned}
\end{equation}
The induced dipole moment along the laser polarization direction from the permanent dipole moment $\mu(t)$ and the electric dipole polarizability $\alpha(t)$ can be calculated via the density Matrix,
\begin{equation}\label{eq19}
\begin{aligned}
\mu(t)=\mathrm{Tr}\left[\hat{\rho}(t)\hat{d}_{1}\right];\alpha(t)=\mathrm{Tr}\left[\hat{\rho}(t)\hat{d}_{2}\right].
\end{aligned}
\end{equation}

Thus, the radiation spectra generated by the induced dipole moment can be gotten by
\begin{equation}\label{eq20}
\begin{aligned}
S(\omega)\varpropto\left|\frac{1}{T}\int\ddot{D}(t)e^{-i\omega t}dt\right|^{2}.
\end{aligned}
\end{equation}

Additionally, the phononic spectra from the vibrating molecule can be calculated using the Fourier transformation of second derivative of the nuclear displacement with respect to time. Note: While the term of `phonon' does not typically apply to molecular vibrations, by viewing phononic comb as a universal vibratory phenomenon, we used the term `phononic spectra' for also describing the dynamics of molecules. 

We only consider the phononic spectra from harmonic vibration, thus define the nuclear displacement operator $\hat{R}$ along the molecule axis. The matrix elements of the nuclear displacement operator $\hat{R}$ are given by
\begin{equation}\label{eq21}
\begin{aligned}
R_{\nu Jm,\nu'J'm'}&=\langle\nu Jm|(R-R_{e})\cos\theta|\nu'J'm'\rangle.
\end{aligned}
\end{equation}
Thus, the evolution of nuclear displacement out of the equilibrium position driven by the pump laser is expressed as
\begin{equation}\label{eq22}
\begin{aligned}
R(t)=\mathrm{Tr}\left[\hat{\rho}(t)\hat{R}\right].
\end{aligned}
\end{equation}
The phononic frequency comb can be described as
\begin{equation}\label{eq23}
\begin{aligned}
P(\omega)\varpropto\left|\frac{1}{T}\int\ddot{R}(t)e^{-i\omega t}dt\right|^{2}.
\end{aligned}
\end{equation}
\subsection{C. Numerical considerations}
In the calculation, we study the dynamics of rovibrational excitation using the density matrix formalism including ten vibrational levels ($\nu=0\sim9$) and forty rotational states ($J=0\sim39$) with the projection $m=-J\sim J$ without loss of generality. The Liouville-von Neumann equation Eq. (\ref{eq10}) is numerically solved using the fourth-order Runge-Kutta method with time step $\Delta t=0.2$ fs to obtain the dynamics of CO molecule within the driving laser.

The integrals of Eqs. (\ref{eq13}) and (\ref{eq14}) used to calculate the transition matrix elements can be decomposed into integrals of radial wave function $\psi_{\nu J}(R)$ over the internuclear distance $R$ and integrals of spherical harmonics $\mathrm{Y}_{J m}(\theta,\phi)$ over angle $\theta$ and $\phi$. Here, the integral step of internuclear distance is set to be $\Delta R=0.00342$ a.u. and the integral range for the internuclear distance is from $R_{min}=1.5$ a.u. to $R_{max}=5$ a.u., which is sufficient to cover the radial distribution of the eigen wavefunctions under our consideration. The integrations of spherical harmonics $\mathrm{Y}_{J m}(\theta,\phi)$ over angle $\theta$ and $\phi$ are calculated by the Wigner 3-j symbols \cite{Zare1988}.

The radiation spectra and photonic spectra are calculated by the Eq. (\ref{eq20}) and Eq. (\ref{eq23}) with integral time of 250 ps including the signal within and without the pump pulse, from $t=-1$ ps to 249 ps to define the peak time of pulse at $t=0$.

\section{III. Results and Discussion}

We study the generation of PFCs $P(\omega)$ and radiation spectra $S(\omega)$ from CO molecules by numerically solving the Liouville-von Neumann equation of the density matrix, using the initial population distribution based on Boltzmann distribution. To efficiently drive the rovibrational excitation in CO molecules, we employ a resonant optical excitation approach. By tuning to the pump wavelength from 4.672 $\mathrm{\mu}$m to 9.344 $\mathrm{\mu}$m, we simulate the dynamics of rovibrational excitation of CO molecule, focusing on the one-photon and two-photon resonance excitation corresponding to the interaction of permanent electric dipole moment and electric dipole polarizability, respectively.

\subsection{A. One-photon resonance}
\begin{figure}
	\includegraphics*[width=3.3in]{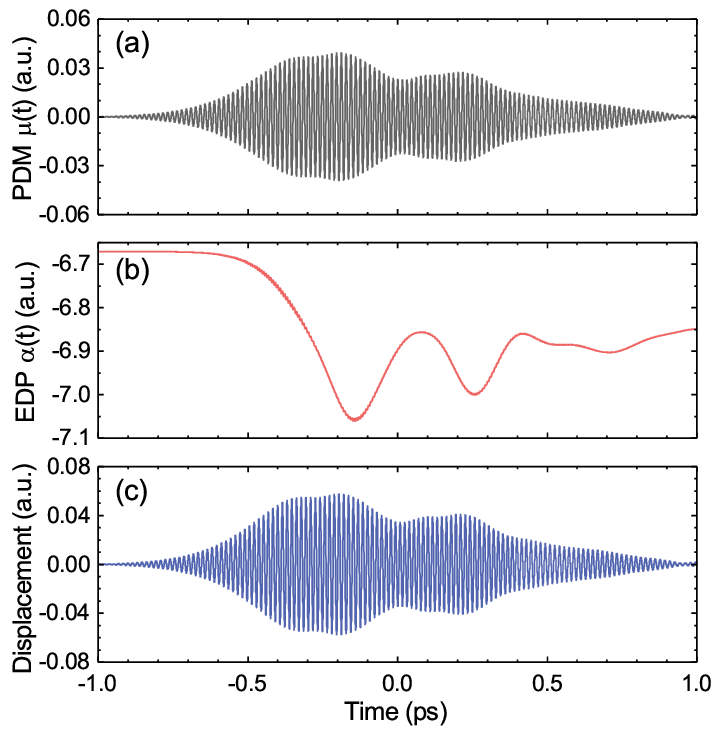}
	\setlength{\abovecaptionskip}{0pt}
	\setlength{\belowcaptionskip}{5pt}
	\caption{
		Dynamics of (a) the electric dipole moment $\mu(t)$, (b) the electric dipole polarizability $\alpha(t)$ and (c) the internuclear displacement out of equilibrium position $R(t)$. Here, the $\nu=0$ to $\nu=1$ transition is in the single-photon resonance with pump laser at the corresponding wavelengths of 4672.1 nm. The laser intensity is 5$\times$10$^{13}$ W/cm$^2$, with pulse duration of 500 fs.
	}\label{f4}
\end{figure}
First, the central wavelength of the pump laser pulse is set to be 4.672 $\mathrm{\mu}$m (wavenumber 2140.4 cm$^{-1}$), corresponding to the transition wavelength between the vibrational levels of $\nu=0, 1$ in the electronic ground-state of CO molecule.  Figure \ref{f4} illustrates the dynamics of the permanent dipole moment, the electric dipole polarizability and the vibrational motion of CO molecules within the laser field. As depicted in Fig. \ref{f4}(a), the electric dipole moment begins to oscillate in the laser field, stimulating the molecule to transition to excited vibrational states. Moreover, it can be known from  Fig. \ref{f4}(b) that the electric dipole polarizability is modulated slowly in the laser field, accompanied by rapid oscillations of small amplitude. Figure \ref{f4}(c) demonstrates that the CO molecule driven by a pumped laser vibrates in the range of -0.06 a.u. to 0.06 a.u.. Meanwhile, the slow modulation observed in Figs. \ref{f4}(a-c) is attributed to the Rabi oscillation. A comparison of Figs. \ref{f4}(a), \ref{f4}(b), and \ref{f4}(c) suggests that the harmonic vibrational motion of the CO molecule is primarily driven by the polarization of the permanent dipole moment, with minimal contribution from the electric dipole polarizability.
\begin{figure}
\includegraphics*[width=3.3 in]{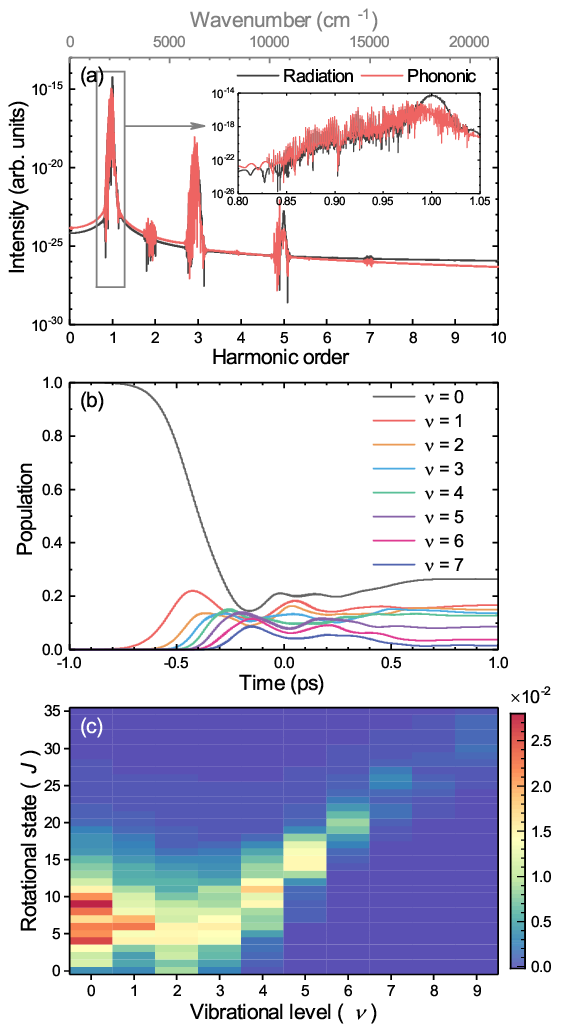}
\setlength{\abovecaptionskip}{0pt}
\setlength{\belowcaptionskip}{0pt}
\caption{
Population dynamics and spectra of CO molecule excited by the driving pulse which is in the single-photon resonance with the $\nu=0$ to $\nu=1$ transition. (a) Radiation spectrum (black line) and phononic spectrum (red line). (b) Population dynamics of the vibrational levels within the laser pulse. (c) The population distribution of rovibrational states at the end of the driving pulse.
}\label{f5}
\end{figure}

We substitute the time-domain evolution of the permanent dipole moment, electric dipole polarizability and molecular internuclear displacement into Eqs.(\ref{eq20}) and (\ref{eq23}) to calculate the radiation and phononic spectra generated by the molecular vibrational excitation, as shown in Fig. \ref{f5}(a). Figure \ref{f5}(a) manifests both the radiation and phononic spectra, which consist of a series of equidistant frequency peaks, with a maximum harmonic order of seven. The radiation and phononic spectra are primarily composed of odd-order harmonics, which are induced by electric dipole interactions that adhere to the selection rule  $\Delta J=\pm1$ and conservation of parity. Consequently, the generation of even-order harmonics is forbidden in quantum systems with opposite symmetry for the electric dipole interaction. 

Meanwhile, when we magnify spectra to observe their fine structure, as illustrated in Fig. \ref{f5}(a), we can discern that the broadened spectra exhibit many nearly equally spaced spikes. By comparing the spectra within and without the pump pulse, we find that the spikes in the harmonic spectra are generated by the free vibration of the molecule at the end of the pump pulse, induced by the vibrational coherence established by the vibrational transition within the pump laser. The spikes in the $n$-th harmonic originate from the vibrational coherence established by the vibrational transitions $\Delta \nu=n$. The weak signal at even harmonics is induced by the non-resonant and weak single-photon transition with even $\Delta\nu$. Furthermore, the spectral positions of these spikes correspond to the rovibrational transition energies of CO molecules, indicating that the variation in transition energy caused by molecular rotation broadens harmonic spectrum. Therefore, we can confirm that the rovibrational excitation of the molecule produces the spectra, providing an approach to analyze the spectral structure based on the population of rovibrational states in the CO molecule. In the case of single-photon resonance, we simulate the dynamics of the population on vibrational levels during the pump pulse and the population distribution of rovibrational states at the conclusion of the driving pulse, as depicted in Figs. \ref{f5}(b) and (c), respectively.

Figure \ref{f5}(b) indicates that the initial population is almost on the vibrational level $\nu=0$ according to the Boltzmann distribution. When the laser is single-photon resonant with the vibrational levels $\nu=0$ and $\nu=1$ of the electronic ground-state CO molecule, the pump laser is also nearly resonant with the other adjacent vibrational states due to the small anharmonic coefficient of the CO molecule, as shown in Figs. \ref{f2}(b)(c). Thus, higher vibrational levels of CO molecules can be populated through cascaded vibrational transitions with $\Delta\nu=\pm1$. The time at which the maximum of each vibrational population occurs is delayed as the vibrational quantum number increases, which also means that the vibrational transitions with $\Delta \nu=\pm1$ dominate the transfer process of the vibrational population. Simultaneously, under the single-photon resonance excitation, various vibrational populations undergo different Rabi oscillations, contributing to further broadening of harmonics and the excited vibrational states (i.e., $\nu=2, 3,\ldots$) can be readily populated at the end of the pump pulse. During the driving pulse, the redistribution of population caused by Rabi oscillations induces the slow modulation of molecular vibration and polarization over time, as displayed in Fig. \ref{f4}. After the end of the pump pulse, the distribution of vibrational population is determined by the redistribution of Rabi oscillation, which can be controlled by the laser intensity and pulse duration.

The rovibrational population distribution at the end of the pulse is shown in Fig. \ref{f5}(c). Notably, Figure \ref{f5}(c) reveals that the population of higher vibrational levels is always distributed in the higher rotational states after the end of the driving pulse. The origin for such a rovibrational population distribution is that the employed driving pulse spectrum covers only the transition energies of R-branch transition ($\Delta J=+1$) for the higher vibrational levels transition ($\Delta\nu=+1$), as presented in Figs. \ref{f2}(b)(c). Therefore, the transitions with $\Delta\nu=+1$ and $\Delta J=+1$ dominate the population transfer, resulting in higher rotation state being  populated on higher vibrational levels. Additionally, CO molecules can be excited to higher vibrational levels and rotational states using shorter pulse widths to obtain a wider spectrum, which can cover more R-branch transition energies of higher excited states.


\subsection{B. Two-photon resonance}
\par
\begin{figure}
\includegraphics*[width=3.3in]{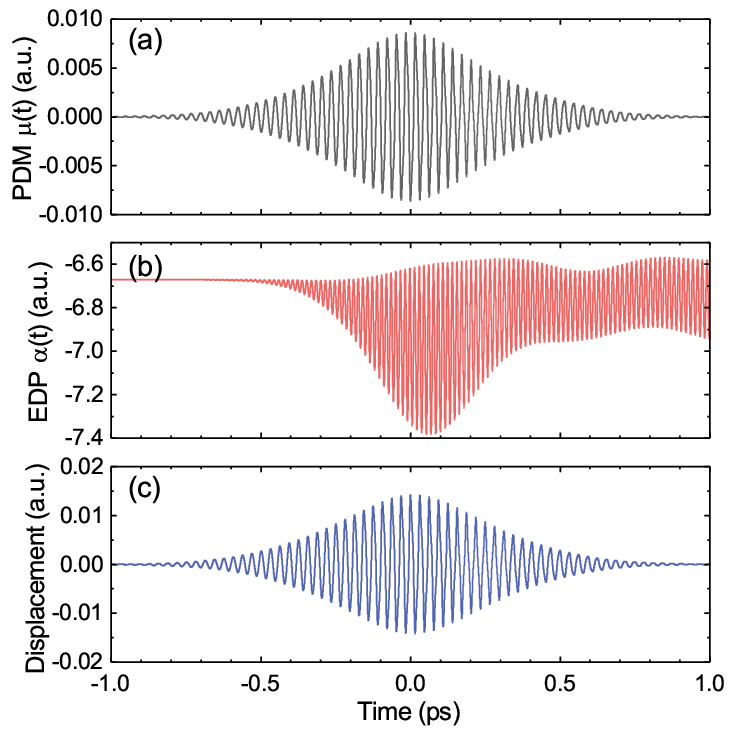}
\setlength{\abovecaptionskip}{0pt}
\setlength{\belowcaptionskip}{5pt}
\caption{
Dynamics of (a) the permanent dipole moment $\mu(t)$, (b) the electric dipole polarizability $\alpha(t)$ and (c) the nuclear displacement out of equilibrium position $R(t)$ under the condition of the two-photon resonance between the $\nu=0$ to $\nu=1$ transition and driving pulse through shifting the laser wavelength to 9.344 $\mathrm{\mu}$m, while keeping other laser parameters constant.
}\label{f6}
\end{figure}

\begin{figure}
\includegraphics*[width=3.3 in]{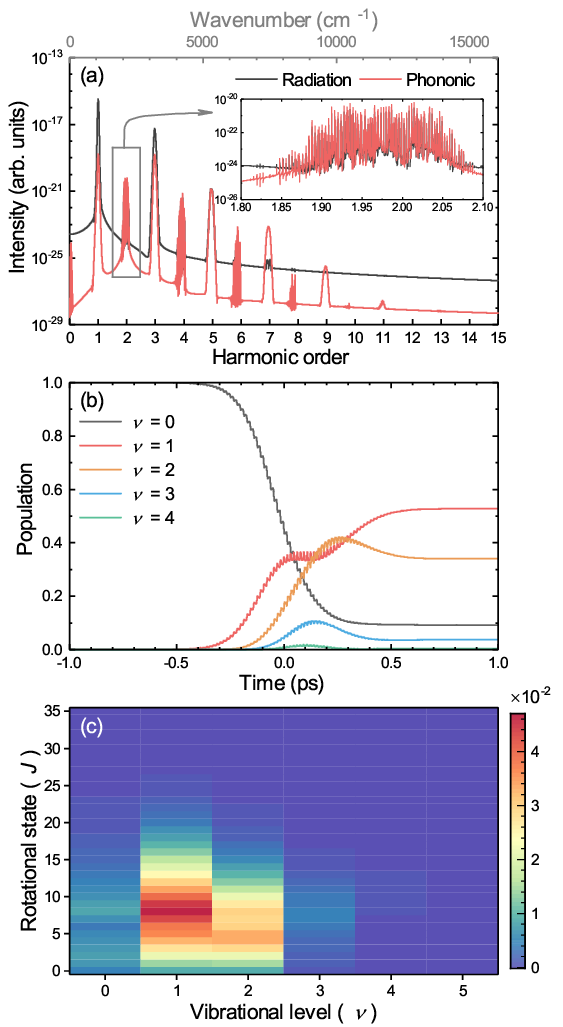}
\setlength{\abovecaptionskip}{0pt}
\setlength{\belowcaptionskip}{5pt}
\caption{
Population dynamics and spectra of CO molecule, when the driving pulse is in two-photon resonance with the vobrational transition from $\nu=0$ to $\nu=1$ through shifting the laser wavelength to 9.344 $\mathrm{\mu}$m, keeping other laser parameters consistent. (a) Radiation spectrum (black line) and phononic spectrum (red line). (b) Population dynamics of the vibrational levels within the laser pulse. (c) The population distribution of rovibrational states at the end of the driving pulse.
}\label{f7}
\end{figure}

 In the following, we numerically solve Eq. (\ref{eq10}) with the two-photon resonance to comparatively investigate the characteristics of PFCs created under the circumstances of different optical resonances and to obtain PFCs of various spectral structures. For the two-photon resonance, we select a laser wavelength of 9.344 $\mathrm{\mu}$m (the wavenumber 1070.2 cm$^{-1}$). The results of the numerical simulation of two-photon resonance are illustrated in Fig. \ref{f6} and Fig. \ref{f7}.

Figure \ref{f6} depicts the temporal evolution of the electric dipole moment, the electric dipole polarizability, and the nuclear displacement out of the equilibrium position, respectively. During the laser pulse, Fig. \ref{f6} shows that the oscillations of the permanent dipole moment and internuclear distance follow that of the laser electric field. The nuclear displacement out of equilibrium position oscillates between -0.015 a.u. and 0.015 a.u., while the electric dipole polarizability undergoes rapid oscillation. However, after the interaction with the pump pulse, the oscillations of the electric dipole moment and internuclear distance fade away, whereas the electric dipole polarizability still maintains a large amplitude oscillation. This indicates that the vibration of CO molecules was weakened, but the rotation persists. Considering the large detuning of single-photon resonance and the two-photon resonance condition, i.e., $\omega_{\nu}=2\omega_{l}$, we deduce that the polarization of the permanent dipole moment does not significantly excite the population to the excited vibrational level. In contrast, the electric dipole polarizability plays a crucial role in vibrational excitation under two-photon resonance conditions.

In Fig. \ref{f7}, we also exhibit the calculated spectra, the population dynamics of vibrational levels, and the rovibrational population distribution under the two-photon resonance condition, analogous to Fig. \ref{f5}.

Figure \ref{f7}(a) presents the harmonic spectra under two-photon resonance condition. Compared with the spectra induced by the single-photon resonance, the maximum order of harmonic spectra from the two photon resonance can reach eleven and the even-order harmonic intensities are relatively stronger. It should be noted that even-order harmonics are generated via the even photon resonance driven by the electric dipole polarizability. The interaction between the electric dipole polarization and pump laser plays a dominant role in the two-photon resonance, enhancing the even-order harmonics. In addition, unlike the spectra generated by the single-photon resonance, the spikes induced by the molecular free vibration after the pump pulse appear only in the even-order harmonics, as shown in the inset of Fig.\ref{f7}(a). Because the frequencies of these spikes correspond to the intrinsic energy intervals between rovibrational states, approximately equal to $2n\omega_{l}$ in the case of two-photon resonance. Thus, the molecular vibration excitation driven by the two-photon resonance can produce the spectra with smooth odd-order harmonics and even harmonics containing fine spikes generated within and without the pulse, respectively. 

Furthermore, we notice that the third, fifth, seventh harmonic get enhanced in spectra, because the excited vibration levels ($\nu=1,2,3$) are populated, as illustrated in Fig. \ref{f7}(b). Specially, driven by the polarization of the permanent dipole moment, the molecules on excited vibration levels can absorb an odd number of photon and subsequently transition to a lower vibration level, thereby radiating an photon with energy of $n\omega_{l}+\omega_{\nu}\approx m\omega_{l}$, where both $n$ and $m$ are odd, due to the case of two-photon resonance. Therefore, the occupied vibrational excited level with vibrational quantum number $\nu>0$ can enhance harmonics with the order of $2\nu+1$, $2\nu-1$, $\cdots$, i.e., the participation of the excited vibration levels should be responsible for the enhanced odd order harmonic emission in the case. Besides, it can be inferred from Fig. \ref{f7}(b) that CO molecule is excited to the high excited vibration levels through the even-photon transition process induced by the polarization of electric dipole polarizability within the laser field, considering the case of two-photon resonance. The population of the excited vibrational level $\nu=2$ arises from the strong two-photon transition from $\nu=1$ to $\nu=2$ and the weak four-photon transition from $\nu=0$ to $\nu=2$. Moreover, the small population on the higher vibrational levels $\nu=3, 4$ is ascribed to polarization effects and weaker two-photon transitions from adjacent vibrational levels.

The population distribution of rovibrational states at the end of the pump laser under the two-photon resonance condition is depicted in Fig. \ref{f7}(c). The results indicate that CO molecules are excited to the high vibrational levels with $\nu<4$ via the two-photon resonance. In particular, the low energy rotational states on the vibrational levels are predominantly populated, differing from the case of single-photon resonance. The rovibrational excitation is stimulated by the even photon transition induced by the electric dipole polarizability and the spectra of the employed pulse primarily covers the Q-branch transition with $\Delta J=0$ and $\Delta\nu=\pm1$ under the two-photon resonance condition, as shown in Fig. \ref{f2}(c). This indicates that the driving pulse does not significantly excite low rotational states to higher ones. Consequently, low rotational states remain largely occupied after the laser pulse ends, consistent with an initial rotational population distribution following the Boltzmann distribution. Comparing the Fig. \ref{f5}(c) with the Fig. \ref{f7}(c), the rotational population distribution prepared by the two-photon resonance results in a narrower spectrum width and fewer spikes in comparison to that produced under the single-photon resonance condition, as illustrated in the sets of Fig. \ref{f5}(a) and Fig. \ref{f7}(a). This difference arises from the fact that fewer rotational states are occupied in the two-photon resonance case. Hence, we can control the harmonic width and the number of spikes in the harmonics by adjusting the resonance rotational transition channels in the coverage of the pump laser spectrum.

Contrasting the radiation spectra with the phononic spectra from our simulations reveals that the phononic spectra exhibit similar structures to the radiation spectra, allowing for mutual reconstruction. It is important to note that a shorter pulse laser with a broader frequency bandwidth can cover more transition channels and produces a weaker, broader spectrum with higher-order harmonics. Additionally, we can adjust the coverage of the pump spectra to control both the harmonic width and the number of spikes in the harmonics by manipulating the occupancy of low or high-energy rotational states, as well as the intensities of single-photon and two-photon transitions. Finally, by appropriately tuning the pump wavelength, intensity and pulse width, we can shape the phononic spectra arbitrarily to produce novel phononic combs.

\section{IV. Summary}
In summary, we demonstrated the excitation of PFCs in laser-driven molecules via solving the Liouville-von
Neumann equation of density matrix characterizing the molecular rovibrational motion in a laser field under the circumstance of single and two photon resonance. Moreover, it is worth mentioning that the polarization of permanent dipole moment and electric dipole polarizability play a significant role in the odd-number photon and the even-number photon transition, respectively. Furthermore, our results suggest that the different types of multiphoton resonance enhance the corresponding order and can be utilized to transform the spectral characteristic. In our research, we neglect the electron response including the effects of excitation and polarization. However, the laser-induced electronic polarization may influence the nuclear vibration, which will be discussed in our future work.

\section{V. Outlook}
This paper presents a basic pathway for generating phononic combs in molecules. To demonstrate this phenomenon, we considered a molecule conceptualized as a linear vibrator, whose dynamics modelled by a rovibrational density matrix theory. However, in practice, more complex configurations of molecules that involve rotation-vibration couplings \cite{Pekeris1934}, substrate-molecule couplings \cite{Backus2005}, electron-phonon couplings \cite{Devos1998} and laser-induced alignment \cite{Stapelfeldt2003} could be harnessed for phononic comb studies and applications. While the experimental works of phononic combs have so far been bound to mechanical resonators, this paper motivates the possible extension of combs to molecular systems. The implementation of phononic combs in molecules will not only push these combs into infrared radiation (IR) or far-infrared radiation (FIR) regime by harnessing THz molecular vibrations, but also could lead to new prospects in molecular science for PFCs generation.

\section{Acknowledgments}
This work is supported by the Major Research Plan of NSF (Grant No. 91850201), China National Key R\&D  Program (Grant No. 2019YFA0307703) and the National Natural Science Foundation of China (Grant Nos. 12074063 and 11904400).

\bibliography{ref}

\begin{thebibliography}{32}%
\makeatletter
\providecommand \@ifxundefined [1]{%
 \@ifx{#1\undefined}
}%
\providecommand \@ifnum [1]{%
 \ifnum #1\expandafter \@firstoftwo
 \else \expandafter \@secondoftwo
 \fi
}%
\providecommand \@ifx [1]{%
 \ifx #1\expandafter \@firstoftwo
 \else \expandafter \@secondoftwo
 \fi
}%
\providecommand \natexlab [1]{#1}%
\providecommand \enquote  [1]{``#1''}%
\providecommand \bibnamefont  [1]{#1}%
\providecommand \bibfnamefont [1]{#1}%
\providecommand \citenamefont [1]{#1}%
\providecommand \href@noop [0]{\@secondoftwo}%
\providecommand \href [0]{\begingroup \@sanitize@url \@href}%
\providecommand \@href[1]{\@@startlink{#1}\@@href}%
\providecommand \@@href[1]{\endgroup#1\@@endlink}%
\providecommand \@sanitize@url [0]{\catcode `\\12\catcode `\$12\catcode `\&12\catcode `\#12\catcode `\^12\catcode `\_12\catcode `\%12\relax}%
\providecommand \@@startlink[1]{}%
\providecommand \@@endlink[0]{}%
\providecommand \url  [0]{\begingroup\@sanitize@url \@url }%
\providecommand \@url [1]{\endgroup\@href {#1}{\urlprefix }}%
\providecommand \urlprefix  [0]{URL }%
\providecommand \Eprint [0]{\href }%
\providecommand \doibase [0]{http://dx.doi.org/}%
\providecommand \selectlanguage [0]{\@gobble}%
\providecommand \bibinfo  [0]{\@secondoftwo}%
\providecommand \bibfield  [0]{\@secondoftwo}%
\providecommand \translation [1]{[#1]}%
\providecommand \BibitemOpen [0]{}%
\providecommand \bibitemStop [0]{}%
\providecommand \bibitemNoStop [0]{.\EOS\space}%
\providecommand \EOS [0]{\spacefactor3000\relax}%
\providecommand \BibitemShut  [1]{\csname bibitem#1\endcsname}%
\let\auto@bib@innerbib\@empty
\bibitem [{\citenamefont {Ganesan}\ \emph {et~al.}(2017)\citenamefont {Ganesan}, \citenamefont {Do},\ and\ \citenamefont {Seshia}}]{GanesanPRL2017}%
  \BibitemOpen
  \bibfield  {author} {\bibinfo {author} {\bibfnamefont {A.}~\bibnamefont {Ganesan}}, \bibinfo {author} {\bibfnamefont {C.}~\bibnamefont {Do}}, \ and\ \bibinfo {author} {\bibfnamefont {A.}~\bibnamefont {Seshia}},\ }\href {\doibase 10.1103/PhysRevLett.118.033903} {\bibfield  {journal} {\bibinfo  {journal} {Phys. Rev. Lett.}\ }\textbf {\bibinfo {volume} {118}},\ \bibinfo {pages} {033903} (\bibinfo {year} {2017})}\BibitemShut {NoStop}%
\bibitem [{\citenamefont {Ganesan}\ \emph {et~al.}(2018{\natexlab{a}})\citenamefont {Ganesan}, \citenamefont {Do},\ and\ \citenamefont {Seshia}}]{GanesanAPL2018}%
  \BibitemOpen
  \bibfield  {author} {\bibinfo {author} {\bibfnamefont {A.}~\bibnamefont {Ganesan}}, \bibinfo {author} {\bibfnamefont {C.}~\bibnamefont {Do}}, \ and\ \bibinfo {author} {\bibfnamefont {A.}~\bibnamefont {Seshia}},\ }\href {\doibase 10.1063/1.5003133} {\bibfield  {journal} {\bibinfo  {journal} {Appl. Phys. Lett.}\ }\textbf {\bibinfo {volume} {112}},\ \bibinfo {pages} {021906} (\bibinfo {year} {2018}{\natexlab{a}})}\BibitemShut {NoStop}%
\bibitem [{\citenamefont {Ganesan}\ \emph {et~al.}(2018{\natexlab{b}})\citenamefont {Ganesan}, \citenamefont {Do},\ and\ \citenamefont {Seshia}}]{GanesanPRB2018}%
  \BibitemOpen
  \bibfield  {author} {\bibinfo {author} {\bibfnamefont {A.}~\bibnamefont {Ganesan}}, \bibinfo {author} {\bibfnamefont {C.}~\bibnamefont {Do}}, \ and\ \bibinfo {author} {\bibfnamefont {A.}~\bibnamefont {Seshia}},\ }\href {\doibase 10.1103/PhysRevB.97.014302} {\bibfield  {journal} {\bibinfo  {journal} {Phys. Rev. B}\ }\textbf {\bibinfo {volume} {97}},\ \bibinfo {pages} {014302} (\bibinfo {year} {2018}{\natexlab{b}})}\BibitemShut {NoStop}%
\bibitem [{\citenamefont {Czaplewski}\ \emph {et~al.}(2018)\citenamefont {Czaplewski}, \citenamefont {Chen}, \citenamefont {Lopez}, \citenamefont {Shoshani}, \citenamefont {Eriksson}, \citenamefont {Strachan},\ and\ \citenamefont {Shaw}}]{Czaplewski2018}%
  \BibitemOpen
  \bibfield  {author} {\bibinfo {author} {\bibfnamefont {D.~A.}\ \bibnamefont {Czaplewski}}, \bibinfo {author} {\bibfnamefont {C.}~\bibnamefont {Chen}}, \bibinfo {author} {\bibfnamefont {D.}~\bibnamefont {Lopez}}, \bibinfo {author} {\bibfnamefont {O.}~\bibnamefont {Shoshani}}, \bibinfo {author} {\bibfnamefont {A.~M.}\ \bibnamefont {Eriksson}}, \bibinfo {author} {\bibfnamefont {S.}~\bibnamefont {Strachan}}, \ and\ \bibinfo {author} {\bibfnamefont {S.~W.}\ \bibnamefont {Shaw}},\ }\href {\doibase 10.1103/PhysRevLett.121.244302} {\bibfield  {journal} {\bibinfo  {journal} {Phys. Rev. Lett.}\ }\textbf {\bibinfo {volume} {121}},\ \bibinfo {pages} {244302} (\bibinfo {year} {2018})}\BibitemShut {NoStop}%
\bibitem [{\citenamefont {Goryachev}\ \emph {et~al.}(2020)\citenamefont {Goryachev}, \citenamefont {Galliou},\ and\ \citenamefont {Tobar}}]{Goryachev2020}%
  \BibitemOpen
  \bibfield  {author} {\bibinfo {author} {\bibfnamefont {M.}~\bibnamefont {Goryachev}}, \bibinfo {author} {\bibfnamefont {S.}~\bibnamefont {Galliou}}, \ and\ \bibinfo {author} {\bibfnamefont {M.~E.}\ \bibnamefont {Tobar}},\ }\href {\doibase 10.1103/PhysRevResearch.2.023035} {\bibfield  {journal} {\bibinfo  {journal} {Phys. Rev. Research}\ }\textbf {\bibinfo {volume} {2}},\ \bibinfo {pages} {023035} (\bibinfo {year} {2020})}\BibitemShut {NoStop}%
\bibitem [{\citenamefont {Singh}\ \emph {et~al.}(2020)\citenamefont {Singh}, \citenamefont {Sarkar}, \citenamefont {Guria}, \citenamefont {Nicholl}, \citenamefont {Chakraborty}, \citenamefont {Bolotin},\ and\ \citenamefont {Ghosh}}]{Singh2020}%
  \BibitemOpen
  \bibfield  {author} {\bibinfo {author} {\bibfnamefont {R.}~\bibnamefont {Singh}}, \bibinfo {author} {\bibfnamefont {A.}~\bibnamefont {Sarkar}}, \bibinfo {author} {\bibfnamefont {C.}~\bibnamefont {Guria}}, \bibinfo {author} {\bibfnamefont {R.~J.}\ \bibnamefont {Nicholl}}, \bibinfo {author} {\bibfnamefont {S.}~\bibnamefont {Chakraborty}}, \bibinfo {author} {\bibfnamefont {K.~I.}\ \bibnamefont {Bolotin}}, \ and\ \bibinfo {author} {\bibfnamefont {S.}~\bibnamefont {Ghosh}},\ }\href {\doibase 10.1021/acs.nanolett.0c01586} {\bibfield  {journal} {\bibinfo  {journal} {Nano Lett.}\ }\textbf {\bibinfo {volume} {20}},\ \bibinfo {pages} {4659} (\bibinfo {year} {2020})}\BibitemShut {NoStop}%
\bibitem [{\citenamefont {Chiout}\ \emph {et~al.}(2021)\citenamefont {Chiout}, \citenamefont {Correia}, \citenamefont {Zhao}, \citenamefont {Johnson}, \citenamefont {Pierucci}, \citenamefont {Oehler}, \citenamefont {Ouerghi},\ and\ \citenamefont {Chaste}}]{chiout2021multi}%
  \BibitemOpen
  \bibfield  {author} {\bibinfo {author} {\bibfnamefont {A.}~\bibnamefont {Chiout}}, \bibinfo {author} {\bibfnamefont {F.}~\bibnamefont {Correia}}, \bibinfo {author} {\bibfnamefont {M.-Q.}\ \bibnamefont {Zhao}}, \bibinfo {author} {\bibfnamefont {A.}~\bibnamefont {Johnson}}, \bibinfo {author} {\bibfnamefont {D.}~\bibnamefont {Pierucci}}, \bibinfo {author} {\bibfnamefont {F.}~\bibnamefont {Oehler}}, \bibinfo {author} {\bibfnamefont {A.}~\bibnamefont {Ouerghi}}, \ and\ \bibinfo {author} {\bibfnamefont {J.}~\bibnamefont {Chaste}},\ }\href {https://doi.org/10.1063/5.0059015} {\bibfield  {journal} {\bibinfo  {journal} {Appl. Phys. Lett.}\ }\textbf {\bibinfo {volume} {119}} (\bibinfo {year} {2021})}\BibitemShut {NoStop}%
\bibitem [{\citenamefont {Nguyen}\ \emph {et~al.}(2021)\citenamefont {Nguyen}, \citenamefont {Maksymov},\ and\ \citenamefont {Suslov}}]{nguyen2021acoustic}%
  \BibitemOpen
  \bibfield  {author} {\bibinfo {author} {\bibfnamefont {B.~Q.~H.}\ \bibnamefont {Nguyen}}, \bibinfo {author} {\bibfnamefont {I.~S.}\ \bibnamefont {Maksymov}}, \ and\ \bibinfo {author} {\bibfnamefont {S.~A.}\ \bibnamefont {Suslov}},\ }\href {https://doi.org/10.1038/s41598-020-79567-6} {\bibfield  {journal} {\bibinfo  {journal} {Sci. Rep.}\ }\textbf {\bibinfo {volume} {11}},\ \bibinfo {pages} {38} (\bibinfo {year} {2021})}\BibitemShut {NoStop}%
\bibitem [{\citenamefont {Qi}\ \emph {et~al.}(2020)\citenamefont {Qi}, \citenamefont {Menyuk}, \citenamefont {Gorman},\ and\ \citenamefont {Ganesan}}]{Qi2020}%
  \BibitemOpen
  \bibfield  {author} {\bibinfo {author} {\bibfnamefont {Z.}~\bibnamefont {Qi}}, \bibinfo {author} {\bibfnamefont {C.~R.}\ \bibnamefont {Menyuk}}, \bibinfo {author} {\bibfnamefont {J.~J.}\ \bibnamefont {Gorman}}, \ and\ \bibinfo {author} {\bibfnamefont {A.}~\bibnamefont {Ganesan}},\ }\href {\doibase 10.1063/5.0025314} {\bibfield  {journal} {\bibinfo  {journal} {Appl. Phys. Lett.}\ }\textbf {\bibinfo {volume} {117}},\ \bibinfo {pages} {183503} (\bibinfo {year} {2020})}\BibitemShut {NoStop}%
\bibitem [{\citenamefont {de~Jong}\ \emph {et~al.}(2023)\citenamefont {de~Jong}, \citenamefont {Ganesan}, \citenamefont {Cupertino}, \citenamefont {Gr{\"o}blacher},\ and\ \citenamefont {Norte}}]{de2023mechanical}%
  \BibitemOpen
  \bibfield  {author} {\bibinfo {author} {\bibfnamefont {M.~H.}\ \bibnamefont {de~Jong}}, \bibinfo {author} {\bibfnamefont {A.}~\bibnamefont {Ganesan}}, \bibinfo {author} {\bibfnamefont {A.}~\bibnamefont {Cupertino}}, \bibinfo {author} {\bibfnamefont {S.}~\bibnamefont {Gr{\"o}blacher}}, \ and\ \bibinfo {author} {\bibfnamefont {R.~A.}\ \bibnamefont {Norte}},\ }\href {https://doi.org/10.1038/s41467-023-36953-8} {\bibfield  {journal} {\bibinfo  {journal} {Nat. Commun.}\ }\textbf {\bibinfo {volume} {14}},\ \bibinfo {pages} {1458} (\bibinfo {year} {2023})}\BibitemShut {NoStop}%
\bibitem [{\citenamefont {Aubanel}\ \emph {et~al.}(1994)\citenamefont {Aubanel}, \citenamefont {Zuo},\ and\ \citenamefont {Bandrauk}}]{Aubanel1994}%
  \BibitemOpen
  \bibfield  {author} {\bibinfo {author} {\bibfnamefont {E.~E.}\ \bibnamefont {Aubanel}}, \bibinfo {author} {\bibfnamefont {T.}~\bibnamefont {Zuo}}, \ and\ \bibinfo {author} {\bibfnamefont {A.~D.}\ \bibnamefont {Bandrauk}},\ }\href {\doibase 10.1103/PhysRevA.49.3776} {\bibfield  {journal} {\bibinfo  {journal} {Phys. Rev. A}\ }\textbf {\bibinfo {volume} {49}},\ \bibinfo {pages} {3776} (\bibinfo {year} {1994})}\BibitemShut {NoStop}%
\bibitem [{\citenamefont {Friedrich}\ and\ \citenamefont {Herschbach}(1995)}]{Friedrich1995}%
  \BibitemOpen
  \bibfield  {author} {\bibinfo {author} {\bibfnamefont {B.}~\bibnamefont {Friedrich}}\ and\ \bibinfo {author} {\bibfnamefont {D.}~\bibnamefont {Herschbach}},\ }\href {\doibase 10.1103/PhysRevLett.74.4623} {\bibfield  {journal} {\bibinfo  {journal} {Phys. Rev. Lett.}\ }\textbf {\bibinfo {volume} {74}},\ \bibinfo {pages} {4623} (\bibinfo {year} {1995})}\BibitemShut {NoStop}%
\bibitem [{\citenamefont {Hoang}\ \emph {et~al.}(2017)\citenamefont {Hoang}, \citenamefont {Zhao}, \citenamefont {Le},\ and\ \citenamefont {Le}}]{Hoang2017}%
  \BibitemOpen
  \bibfield  {author} {\bibinfo {author} {\bibfnamefont {V.-H.}\ \bibnamefont {Hoang}}, \bibinfo {author} {\bibfnamefont {S.-F.}\ \bibnamefont {Zhao}}, \bibinfo {author} {\bibfnamefont {V.-H.}\ \bibnamefont {Le}}, \ and\ \bibinfo {author} {\bibfnamefont {A.-T.}\ \bibnamefont {Le}},\ }\href {\doibase 10.1103/PhysRevA.95.023407} {\bibfield  {journal} {\bibinfo  {journal} {Phys. Rev. A}\ }\textbf {\bibinfo {volume} {95}},\ \bibinfo {pages} {023407} (\bibinfo {year} {2017})}\BibitemShut {NoStop}%
\bibitem [{\citenamefont {Blum}(2012)}]{Blum2012}%
  \BibitemOpen
  \bibfield  {author} {\bibinfo {author} {\bibfnamefont {K.}~\bibnamefont {Blum}},\ }\href {https://link.springer.com/book/10.1007/978-3-642-20561-3} {\emph {\bibinfo {title} {Density matrix theory and applications}}}\ (\bibinfo  {publisher} {Springer Berlin, Heidelberg},\ \bibinfo {year} {2012})\BibitemShut {NoStop}%
\bibitem [{\citenamefont {Seideman}\ and\ \citenamefont {Hamilton}(2005)}]{Seideman2005}%
  \BibitemOpen
  \bibfield  {author} {\bibinfo {author} {\bibfnamefont {T.}~\bibnamefont {Seideman}}\ and\ \bibinfo {author} {\bibfnamefont {E.}~\bibnamefont {Hamilton}},\ }\href {https://www.sciencedirect.com/science/article/pii/S1049250X05520068} {\ \bibinfo {series} {Advances In AMO Physics},\ \textbf {\bibinfo {volume} {52}},\ \bibinfo {pages} {289} (\bibinfo {year} {2005})}\BibitemShut {NoStop}%
\bibitem [{\citenamefont {Bransden}\ and\ \citenamefont {Joachain}(1982)}]{Bransden1982}%
  \BibitemOpen
  \bibfield  {author} {\bibinfo {author} {\bibfnamefont {B.~H.}\ \bibnamefont {Bransden}}\ and\ \bibinfo {author} {\bibfnamefont {C.}~\bibnamefont {Joachain}},\ }\href@noop {} {\emph {\bibinfo {title} {Physics of atoms and molecules}}}\ (\bibinfo  {publisher} {John Wiley \& Sons, New York},\ \bibinfo {year} {1982})\BibitemShut {NoStop}%
\bibitem [{\citenamefont {Brand}\ and\ \citenamefont {Speakman}(1960)}]{Brand1960}%
  \BibitemOpen
  \bibfield  {author} {\bibinfo {author} {\bibfnamefont {J.~C.~D.}\ \bibnamefont {Brand}}\ and\ \bibinfo {author} {\bibfnamefont {J.~C.}\ \bibnamefont {Speakman}},\ }\href@noop {} {\emph {\bibinfo {title} {Molecular Structure: The Physical Approach}}}\ (\bibinfo  {publisher} {John Wiley \& Sons, New York},\ \bibinfo {year} {1960})\BibitemShut {NoStop}%
\bibitem [{\citenamefont {Borges}\ \emph {et~al.}(2001)\citenamefont {Borges}, \citenamefont {Caridade},\ and\ \citenamefont {Varandas}}]{Borges2001}%
  \BibitemOpen
  \bibfield  {author} {\bibinfo {author} {\bibfnamefont {I.}~\bibnamefont {Borges}}, \bibinfo {author} {\bibfnamefont {P.~J. S.~B.}\ \bibnamefont {Caridade}}, \ and\ \bibinfo {author} {\bibfnamefont {A.~J.~C.}\ \bibnamefont {Varandas}},\ }\href {\doibase https://doi.org/10.1006/jmsp.2001.8402} {\bibfield  {journal} {\bibinfo  {journal} {J. Microelectromech. S.}\ }\textbf {\bibinfo {volume} {209}},\ \bibinfo {pages} {24} (\bibinfo {year} {2001})}\BibitemShut {NoStop}%
\bibitem [{\citenamefont {Langhoff}\ and\ \citenamefont {Bauschlicher}(1995)}]{Langhoff1995}%
  \BibitemOpen
  \bibfield  {author} {\bibinfo {author} {\bibfnamefont {S.~R.}\ \bibnamefont {Langhoff}}\ and\ \bibinfo {author} {\bibfnamefont {C.~W.}\ \bibnamefont {Bauschlicher}},\ }\href {\doibase 10.1063/1.469247} {\bibfield  {journal} {\bibinfo  {journal} {J. Chem. Phys.}\ }\textbf {\bibinfo {volume} {102}},\ \bibinfo {pages} {5220} (\bibinfo {year} {1995})}\BibitemShut {NoStop}%
\bibitem [{\citenamefont {Maroulis}(1996)}]{Maroulis1996}%
  \BibitemOpen
  \bibfield  {author} {\bibinfo {author} {\bibfnamefont {G.}~\bibnamefont {Maroulis}},\ }\href {\doibase 10.1021/jp960412n} {\bibfield  {journal} {\bibinfo  {journal} {J. Phys. Chem.}\ }\textbf {\bibinfo {volume} {100}},\ \bibinfo {pages} {13466} (\bibinfo {year} {1996})}\BibitemShut {NoStop}%
\bibitem [{\citenamefont {Sunil}\ and\ \citenamefont {Jordan}(1988)}]{Sunil1988}%
  \BibitemOpen
  \bibfield  {author} {\bibinfo {author} {\bibfnamefont {K.}~\bibnamefont {Sunil}}\ and\ \bibinfo {author} {\bibfnamefont {K.}~\bibnamefont {Jordan}},\ }\href {\doibase https://doi.org/10.1016/0009-2614(88)80194-5} {\bibfield  {journal} {\bibinfo  {journal} {Chem. Phys. Lett.}\ }\textbf {\bibinfo {volume} {145}},\ \bibinfo {pages} {377} (\bibinfo {year} {1988})}\BibitemShut {NoStop}%
\bibitem [{\citenamefont {Lefebvre‐Brion}\ \emph {et~al.}(1961)\citenamefont {Lefebvre‐Brion}, \citenamefont {Moser},\ and\ \citenamefont {Nesbet}}]{Helene1961}%
  \BibitemOpen
  \bibfield  {author} {\bibinfo {author} {\bibfnamefont {H.}~\bibnamefont {Lefebvre‐Brion}}, \bibinfo {author} {\bibfnamefont {C.~M.}\ \bibnamefont {Moser}}, \ and\ \bibinfo {author} {\bibfnamefont {R.~K.}\ \bibnamefont {Nesbet}},\ }\href {\doibase 10.1063/1.1731799} {\bibfield  {journal} {\bibinfo  {journal} {J. Chem. Phys.}\ }\textbf {\bibinfo {volume} {34}},\ \bibinfo {pages} {1950} (\bibinfo {year} {1961})}\BibitemShut {NoStop}%
\bibitem [{\citenamefont {Pinkham}\ and\ \citenamefont {Jones}(2005)}]{Pinkham2005}%
  \BibitemOpen
  \bibfield  {author} {\bibinfo {author} {\bibfnamefont {D.}~\bibnamefont {Pinkham}}\ and\ \bibinfo {author} {\bibfnamefont {R.~R.}\ \bibnamefont {Jones}},\ }\href {\doibase 10.1103/PhysRevA.72.023418} {\bibfield  {journal} {\bibinfo  {journal} {Phys. Rev. A}\ }\textbf {\bibinfo {volume} {72}},\ \bibinfo {pages} {023418} (\bibinfo {year} {2005})}\BibitemShut {NoStop}%
\bibitem [{\citenamefont {Zhang}\ \emph {et~al.}(2020)\citenamefont {Zhang}, \citenamefont {L\"otstedt},\ and\ \citenamefont {Yamanouchi}}]{zhangyouyuan2020}%
  \BibitemOpen
  \bibfield  {author} {\bibinfo {author} {\bibfnamefont {Y.}~\bibnamefont {Zhang}}, \bibinfo {author} {\bibfnamefont {E.}~\bibnamefont {L\"otstedt}}, \ and\ \bibinfo {author} {\bibfnamefont {K.}~\bibnamefont {Yamanouchi}},\ }\href {\doibase 10.1103/PhysRevA.101.053412} {\bibfield  {journal} {\bibinfo  {journal} {Phys. Rev. A}\ }\textbf {\bibinfo {volume} {101}},\ \bibinfo {pages} {053412} (\bibinfo {year} {2020})}\BibitemShut {NoStop}%
\bibitem [{\citenamefont {Michael~Lewis}\ and\ \citenamefont {Glaser}(2000)}]{Lewis2000}%
  \BibitemOpen
  \bibfield  {author} {\bibinfo {author} {\bibfnamefont {Z.~W.}\ \bibnamefont {Michael~Lewis}}\ and\ \bibinfo {author} {\bibfnamefont {R.}~\bibnamefont {Glaser}},\ }\href {\doibase 10.1021/jp002927r} {\bibfield  {journal} {\bibinfo  {journal} {J. Phys. Chem. A}\ }\textbf {\bibinfo {volume} {104}},\ \bibinfo {pages} {11355} (\bibinfo {year} {2000})}\BibitemShut {NoStop}%
\bibitem [{\citenamefont {Jackson}(1998)}]{Jackson1998}%
  \BibitemOpen
  \bibfield  {author} {\bibinfo {author} {\bibfnamefont {J.~D.}\ \bibnamefont {Jackson}},\ }\href {https://aapt.scitation.org/doi/abs/10.1119/1.19136?journalCode=ajp} {\emph {\bibinfo {title} {Classical Electrodynamics}}}\ (\bibinfo  {publisher} {John Wiley \& Sons, New York},\ \bibinfo {year} {1998})\BibitemShut {NoStop}%
\bibitem [{\citenamefont {McLean}\ and\ \citenamefont {Yoshimine}(1967)}]{McLean1967}%
  \BibitemOpen
  \bibfield  {author} {\bibinfo {author} {\bibfnamefont {A.~D.}\ \bibnamefont {McLean}}\ and\ \bibinfo {author} {\bibfnamefont {M.}~\bibnamefont {Yoshimine}},\ }\href {\doibase 10.1063/1.1712220} {\bibfield  {journal} {\bibinfo  {journal} {The Journal of Chemical Physics}\ }\textbf {\bibinfo {volume} {47}},\ \bibinfo {pages} {1927} (\bibinfo {year} {1967})}\BibitemShut {NoStop}%
\bibitem [{\citenamefont {Zare}(1988)}]{Zare1988}%
  \BibitemOpen
  \bibfield  {author} {\bibinfo {author} {\bibfnamefont {R.~N.}\ \bibnamefont {Zare}},\ }\href@noop {} {\emph {\bibinfo {title} {Angular Momentum: Understanding Spatial Aspects in Chemistry and Physics}}}\ (\bibinfo  {publisher} {John Wiley \& Sons, New York},\ \bibinfo {year} {1988})\BibitemShut {NoStop}%
\bibitem [{\citenamefont {Pekeris}(1934)}]{Pekeris1934}%
  \BibitemOpen
  \bibfield  {author} {\bibinfo {author} {\bibfnamefont {C.~L.}\ \bibnamefont {Pekeris}},\ }\href {\doibase 10.1103/PhysRev.45.98} {\bibfield  {journal} {\bibinfo  {journal} {Phys. Rev.}\ }\textbf {\bibinfo {volume} {45}},\ \bibinfo {pages} {98} (\bibinfo {year} {1934})}\BibitemShut {NoStop}%
\bibitem [{\citenamefont {Backus}\ \emph {et~al.}(2005)\citenamefont {Backus}, \citenamefont {Eichler}, \citenamefont {Kleyn},\ and\ \citenamefont {Bonn}}]{Backus2005}%
  \BibitemOpen
  \bibfield  {author} {\bibinfo {author} {\bibfnamefont {E.~H.~G.}\ \bibnamefont {Backus}}, \bibinfo {author} {\bibfnamefont {A.}~\bibnamefont {Eichler}}, \bibinfo {author} {\bibfnamefont {A.~W.}\ \bibnamefont {Kleyn}}, \ and\ \bibinfo {author} {\bibfnamefont {M.}~\bibnamefont {Bonn}},\ }\href {\doibase 10.1126/science.1120693} {\bibfield  {journal} {\bibinfo  {journal} {Science}\ }\textbf {\bibinfo {volume} {310}},\ \bibinfo {pages} {1790} (\bibinfo {year} {2005})}\BibitemShut {NoStop}%
\bibitem [{\citenamefont {Devos}\ and\ \citenamefont {Lannoo}(1998)}]{Devos1998}%
  \BibitemOpen
  \bibfield  {author} {\bibinfo {author} {\bibfnamefont {A.}~\bibnamefont {Devos}}\ and\ \bibinfo {author} {\bibfnamefont {M.}~\bibnamefont {Lannoo}},\ }\href {\doibase 10.1103/PhysRevB.58.8236} {\bibfield  {journal} {\bibinfo  {journal} {Phys. Rev. B}\ }\textbf {\bibinfo {volume} {58}},\ \bibinfo {pages} {8236} (\bibinfo {year} {1998})}\BibitemShut {NoStop}%
\bibitem [{\citenamefont {Stapelfeldt}\ and\ \citenamefont {Seideman}(2003)}]{Stapelfeldt2003}%
  \BibitemOpen
  \bibfield  {author} {\bibinfo {author} {\bibfnamefont {H.}~\bibnamefont {Stapelfeldt}}\ and\ \bibinfo {author} {\bibfnamefont {T.}~\bibnamefont {Seideman}},\ }\href {\doibase 10.1103/RevModPhys.75.543} {\bibfield  {journal} {\bibinfo  {journal} {Rev. Mod. Phys.}\ }\textbf {\bibinfo {volume} {75}},\ \bibinfo {pages} {543} (\bibinfo {year} {2003})}\BibitemShut {NoStop}%
\end{thebibliography}%

\end{document}